\documentclass[12pt]{article}

\usepackage[all]{xy}
\usepackage{amssymb}
\usepackage{asymptote}
\usepackage{stmaryrd}
\usepackage{hyperref}
\usepackage{listings}

\usepackage[utf8]{inputenc}

\begin{asydef}
import dmct2;
dmct_hr = 0.3;
\end{asydef}

\let\caus=\rightarrow

\title{Causality is Graphically Simple}
\author{Carlos Baquero\\ DI, Universidade do Minho \& INESC TEC}
\date{}

\begin{document}
\maketitle

\paragraph{Introduction} We can say that any computing system executes sequences of actions, with an
action being any relevant change in the state of the system.  For example,
reading a file to memory, modifying the contents of the file in memory, or
writing the new contents to the file, are relevant actions for a text editor.
In a distributed system, actions execute in multiple locations; 
in this context, actions are often named events.  
Examples of events in distributed
systems include sending or receiving  messages, or changing some state in a
node. Not all events are related, but some events can cause and influence how
other, later events, occur. For instance a reply to a received mail message is
influenced by that message, and maybe by other prior messages also received. 

Events in a distributed system can either occur in a close location such as 
different processes running in the same machine, 
at nodes inside a data center, or
geographically spread across the globe, or even at a larger scale in future interplanetary networks. 
Relations of potential cause and effect between events are fundamental
to the design of distributed algorithms, and nowadays few services can claim
not to have some form of distributed algorithm at its core. 

Before we try to make sense of these cause and effect relations, it is
necessary to limit their scope to what can be perceived inside the distributed
system itself \--- we can refer to this as \emph{internal causality}.
Naturally, a distributed system interacts with the rest of the physical world
outside it, and there are also cause and effect relations in that world at
large.  For example, consider a couple planning a night out using a system that
manages reservations for dinners and cinema.  One person reserves the dinner
and calls the other on the phone saying that.  After receiving the phone call,
the second person goes to the system and reserves the cinema.  The reservation management system has no way to know that the first reservation has actually
caused the second one.

This \emph{external causality} cannot be detected by the system, and can only be
approximated by \emph{physical time}.  However, time totally orders all events,
even those unrelated, thus it is no substitute to causality \cite{Sheehy15There}. In this text, we focus on characterizing \emph{internal causality}, the causality that can be tracked by the system. 

\paragraph{Happened-before relation}
This brings us to 1978, when Leslie Lamport defined a partial order, \emph{happened before},
that connects events of a distributed systems that are potentially causally linked \cite{Lamport78}. 
An event $c$ can be the cause of an event $e$, or $c$ happened before $e$, 
if both occur in the same node and $c$ executed first, or, 
being at different nodes, if $e$ could know the occurrence of $c$ thanks to some message 
received from some node that knows about $c$. 
If neither event can know about the other, we say they are concurrent. 

When we say that two events are \emph{potentially} causally linked, we recognize that even if we know an event could have an impact on a causally succeeding one, the semantics of the actual events can make the later one independent from the former. A simple example is captured in Figure \ref{fig:hb:code}, where the assignment of variable \texttt{b} precedes the \texttt{if} statement but does not influence its outcome. 

\begin{figure}
\begin{lstlisting}
a = 5;
b = 5;
if (a > 2) c=2;
\end{lstlisting}
\caption{An assignment event, over variable \texttt{a}, influences the later condition.}
\label{fig:hb:code}
\end{figure}

\begin{figure}[h!]
\begin{center}
\begin{xy}
\xymatrix{
  \mathit{Node\; A(lice)} & \ar@{.}[r] &
  {\bullet} \ar@{->}[r]^<{a_1}_<{\textit{Dinner?}} & \bullet \ar@{->}[rr]^<{a_2} \ar@{->}[rd]
  & & {\bullet} \ar@{.}[rrr]^<{a_3}  & & & \\
  \mathit{Node \; B(ob)} & \ar@{.}[r] & \bullet \ar@{->}[rr]^<{b_1}
&  &
  {\bullet} \ar@{->}[rr]^<{b_2}_<{\textit{Yes, let's do it}} & & \bullet \ar@{.}[rr]^<{b_3} \ar@{->}[rd]
&  & \\
  \mathit{Node\; C(hris)} & \ar@{.}[r] & \bullet \ar@{->}[rrr]^<{c_1}
& & &
  {\bullet} \ar@{->}[rr]^<{c_2}_<{\textit{Bored \ldots}} &
  & {\bullet} \ar@{.}[r]^<{c_3}_<{\textit{Can I join?}} 
& \\
  \mathit{Time} & \ar@{-->}[rrrrrrr] & & & & & & & \\
}
\end{xy} 
\end{center}
\caption{Run in a distributed system with three nodes: happened-before relation.}
\label{fig:hb:example}
\end{figure}
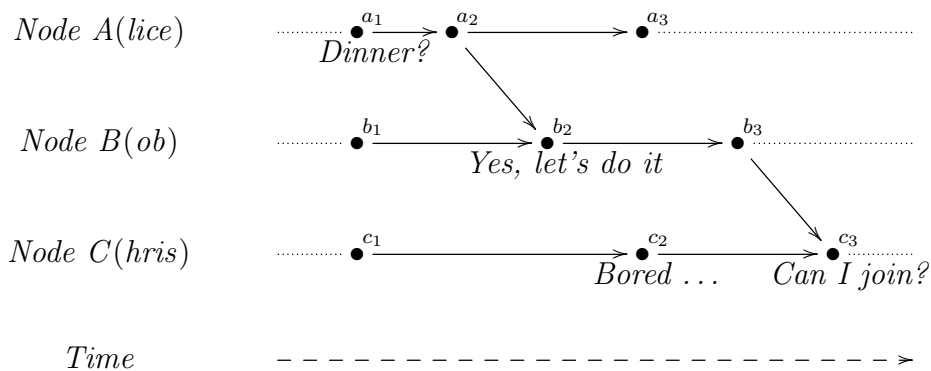

Figure~\ref{fig:hb:example} shows an example of a distributed system. An 
arrow between nodes represents a message sent and delivered.
We can see that both Bob's positive answer to the dinner suggestion by Alice, and Chris later request to join the party, are both influenced by Alice's initial question about plans for dinner. 

Looking at the events in this distributed computation, a simple way to check if
an event $c$ could have caused another event $e$ ($c$ happened before $e$) 
is to find at least one
directed path linking $c$ to $e$. If such a connection is found 
we mark this partial order relation by $c \caus e$ to denote the happened 
before relation or potential causality. For instance we have $a_1 \caus b_2$ and $b_2 \caus c_3$ (and yes, as well $a_1 \caus c_3$, since causality is transitive). 
Events $a_1$ and $c_2$ are concurrent, denoted $a_1 \parallel c_2$, 
because there are no causal paths in either direction. We note $x \parallel y$ 
iff $x \nrightarrow y $ and $y \nrightarrow x$. The fact that Chris was bored didn't influence Alice's question about dinner, nor the other way around. 

We can now recapitulate the three possible relations between two events $x$ and $y$: 
(a) $x$ might have influenced $y$, if $x \caus y$; 
(b) $y$ might have influenced $x$, if $y \caus x$; 
(c) no observable influence among $x$ and $y$, as they occurred concurrently $x \parallel y$.  

\paragraph{Causal Histories}

Causality can be tracked in a very simple way by using \emph{causal histories} \cite{DBLP:journals/dc/SchwarzM94,DBLP:journals/tocs/BirmanJ87}. 
The system can locally assign unique names to each event (e.g. node name and 
local increasing counter) and collect, and transmit, sets of events to capture the known past.   

\begin{figure}[h!]
\begin{center}
\begin{xy}
\xymatrix{
  \mathit{Node\; A} & \ar@{.}[r] &
  {\bullet} \ar@{->}[r]^<{\{\mathbf{a_1}\}} & \bullet \ar@{->}[rr]^<{\{a_1,\mathbf{a_2}\}} \ar@{->}[rd]
  & & {\bullet} \ar@{.}[rrr]^<{\{a_1,a_2,\mathbf{a_3}\}}  & & & \\
  \mathit{Node \; B} & \ar@{.}[r] & \bullet \ar@{->}[rr]^<{\{\mathbf{b_1}\}}
&  &
  {\bullet} \ar@{->}[rr]_<{\{a_1,a_2,b_1,\mathbf{b_2}\}} & & \bullet \ar@{.}[rr]^<{\{a_1,a_2,b_1,b_2,\mathbf{b_3}\}} \ar@{->}[rd]
&  & \\
  \mathit{Node\; C} & \ar@{.}[r] & \bullet \ar@{->}[rrr]_<{\{\mathbf{c_1}\}}
& & &
  {\bullet} \ar@{->}[rr]_<{\{c_1,\mathbf{c_2}\}} &
  & {\bullet} \ar@{.}[r]_<{\{a_1,a_2,b_1,b_2,b_3,c_1,c_2,\mathbf{c_3}\}} 
& \\
  \mathit{Time} & \ar@{-->}[rrrrrrr] & & & & & & & \\
}
\end{xy} 
\end{center}
\caption{Run in a distributed system with three nodes: causal histories.}
\label{fig:ch:example}
\end{figure}
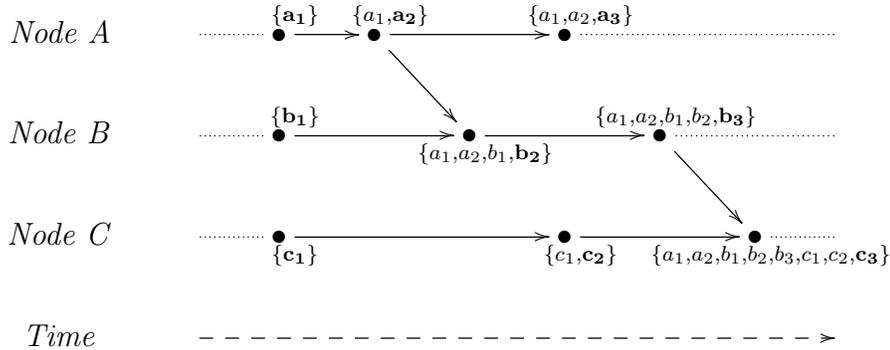

For a new event, the system creates a new unique name and the causal history is
comprised of the union of this name and the causal history of the previous
event in the node. For example, the second event in node $C$ is assigned
name $c_2$ and its causal history is $H_c=\{c_1,\mathbf{c_2}\}$ (shown in Figure
\ref{fig:ch:example}).  When a node sends a message, the causal history of
the send event is sent with the message. On reception, the remote causal causal
history is merged  (by set union) to the local history.  For example, the
delivery of the first message from node $A$ to $B$ merges the remote causal
history, $\{a_1,a_2\}$, with the local history, $\{b_1\}$, and the new unique
name, $b_2$, leading to $\{a_1,a_2,b_1,\mathbf{b_2}\}$. 

Checking causality between two events $x$ and $y$, can be tested simply by set 
inclusion: $x \caus y$ iff $H_x \subset H_y$. 
This follows from the definition of causal histories, where the causal history
of an event will be included in the causal history of the following event.
Even better, if we distinguish the last local event added to the history (denoted in 
\textbf{bold} in the diagram) we can use a simpler test: $x \caus y$ iff $x \in H_y$ -- e.g. 
$a_1 \caus b_2$, since $a_1 \in \{a_1,a_2,b_1,b_2\}$. 
This follows from the fact that a causal history includes all events that (causally) 
precede a given event.

\paragraph{Vector Clocks}
It should be obvious by now, that causal histories work but are not very
compact. We can address this problem by relying on the following observation:
the mechanism of building the causal history implies that if an
event $b_3$ is present in $H_y$, then all preceding events from that same
node, $b_1$ and $b_2$, are also present in $H_y$. Thus, it suffices to
store the most recent event from each node. Causal history
$\{a_1,a_2,b_1,b_2,b_3,c_1,c_2,c_3\}$ is compacted to $\{a \mapsto 2, b \mapsto 3,
c \mapsto 3\}$, or simply a vector $[2,3,3]$. 

Graphically, this can be represented by assigning columns to each source of events and using the column height to depict how many events are known. 

\begin{center}
%
\includegraphics{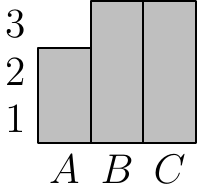}
\end{center}

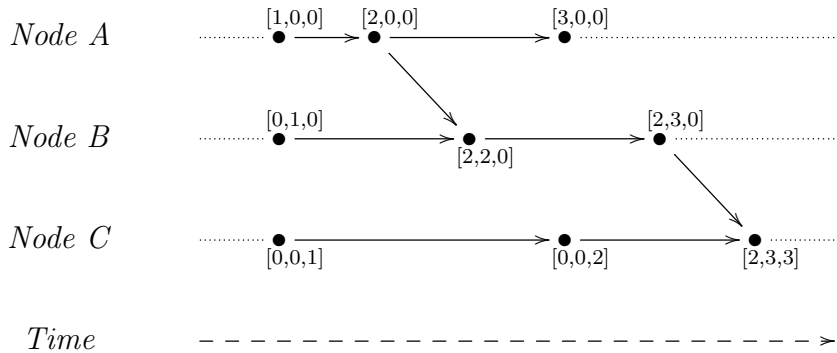
\begin{figure}[h!]
\begin{center}
\begin{xy}
\xymatrix{
  \mathit{Node\; A} & \ar@{.}[r] &
  {\bullet} \ar@{->}[r]^<{[1,0,0]} & \bullet \ar@{->}[rr]^<{[2,0,0]} \ar@{->}[rd]
  & & {\bullet} \ar@{.}[rrr]^<{[3,0,0]}  & & & \\
  \mathit{Node \; B} & \ar@{.}[r] & \bullet \ar@{->}[rr]^<{[0,1,0]}
&  &
  {\bullet} \ar@{->}[rr]_<{[2,2,0]} & & \bullet \ar@{.}[rr]^<{[2,3,0]} \ar@{->}[rd]
&  & \\
  \mathit{Node\; C} & \ar@{.}[r] & \bullet \ar@{->}[rrr]_<{[0,0,1]}
& & &
  {\bullet} \ar@{->}[rr]_<{[0,0,2]} &
  & {\bullet} \ar@{.}[r]_<{[2,3,3]} 
& \\
  \mathit{Time} & \ar@{-->}[rrrrrrr] & & & & & & & \\
}
\end{xy} 
\end{center}
\caption{Run in a distributed system with three nodes: vector clocks.}
\label{fig:vv:example}
\end{figure}

Now, we can translate the rules used with causal histories to the new
compact vector representation.

For verifying that $x \caus y$, we needed to check if $H_x \subset H_y$.
This can be done, verifying for each node, if the unique names contained 
in $H_x$ are also contained in $H_y$ and there is at least one unique name 
in $H_y$ that is not contained in $H_x$. 
This is immediately translated
in checking if each entry in the vector of $x$ is smaller or equal 
to the correspondent entry in the vector of $y$ and one entry is strictly smaller, i.e.,
$\forall i : V_x[i] \leq V_y[i]$ and $\exists j : V_x[j] < V_y[j]$.
This can be stated more compactly by $x \caus y$ iff $V_x < V_y$. Clocks that follow this property are said to characterize causality. 

If we turn to the graphical representation, this translates into testing if one area is strictly covered by the other. For instance checking $[2,3,2] < [2,3,3]$ becomes:

\begin{center}
\begin{tabular}{lll}
\includegraphics{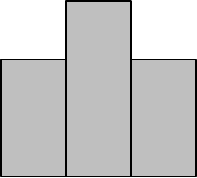}
 & $~~<$ & 
\includegraphics{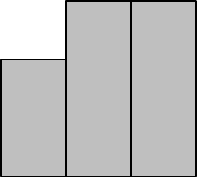}
\end{tabular}
\end{center}

It also becomes intuitive to subtract areas to highlight the additional events:  

\begin{center}
\begin{tabular}{lllll}
\includegraphics{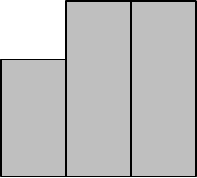}
 & $~~-$ & 
\includegraphics{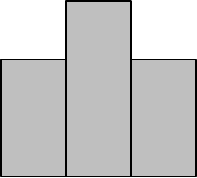}
& $~~=$ &
\includegraphics{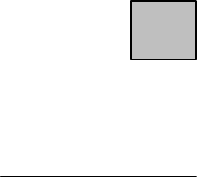}
\end{tabular}
\end{center}

The reader might notice that in general the result of the subtraction is no longer representable by a vector, but can still be captured by causal history notation \--- i.e. $[2,3,3]-[2,3,2]=\{c3\}$. Later in this text we will present efficient encoding notations that can cover some of these cases.  

For a new event, the creation of a new unique name is equivalent to incrementing
the entry in the vector for the node where the event is created.
For example, the third event in node $b$ has vector $[2,3,0]$, 
that corresponds to the creation of event $b_3$ of the causal history.

To capture graphically the requirements for local unique event creation we need to keep track of the identity (column) that is controlled in a give node. This is done by adding an identity layer under the event bar. At node $b$ with vector $[2,2,0]$ we have:

\begin{center}
\includegraphics{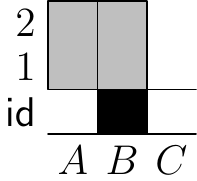}
\end{center}

Having the identity clearly defined, creation of unique events simply translates into an area increase over the controlled identity: 

\begin{center}
\begin{tabular}{lll}
  $ \mathsf{event} ($
\includegraphics{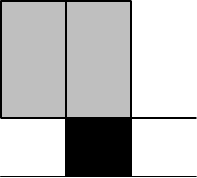}
 $~~~)$ & $~=$ & 
\includegraphics{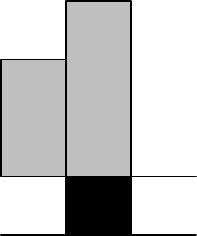}
\end{tabular}
\end{center}

Finally, doing the union of two causal histories, $H_x$ and $H_y$, 
is equivalent to taking the point-wise maximum of the 
correspondent two vectors $V_x$ and $V_y$, i.e., 
$\forall i: V[i] = \mathsf{max}(V_x[i],V_y[i])$. 
The intuition is that, for the unique names generated in each node,
we only need to keep the one with the largest counter. 

When receiving a message, besides merging the causal histories, a new
event is created. The vector representation of these steps can be seen, for
example when the first message from $a$ is received in $b$, where taking the
point-wise maximum leads to $[2,1,0]$ and the new unique name finally leads 
to $[2,2,0]$.

Graphically the point-wise maximum combines the two areas that represent known events, by returning the maximum across all columns. For the above example, the events received from $a$ are combined via $\mathsf{max}$ with those in $b$ by  

\begin{center}
\begin{tabular}{lll}
  $\mathsf{max}($
\includegraphics{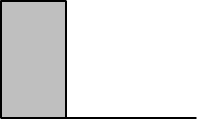}
$,$
\includegraphics{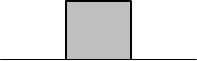}
$~~~)$
 & $~~=$ & 
\includegraphics{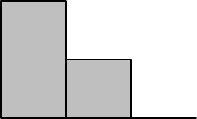}
\end{tabular}
\end{center}

then, all known events are combined with $b$'s identity and a new message reception event is created by $\mathsf{event}$

\begin{center}
\begin{tabular}{lll}
  $ \mathsf{event} ($
\includegraphics{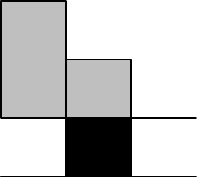}
 $~~~)$ & $~=$ & 
\includegraphics{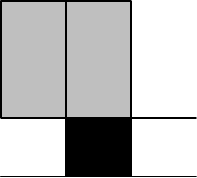}
\end{tabular}
\end{center}

This compact representation is known as \emph{vector clock} and was introduced around
1988 \cite{fidge,mattern}. 
As explained, vector comparison is an immediate translation of 
set inclusion of causal histories, or as simple area inclusion. 
This equivalence is often forgotten in modern descriptions of vector clocks, 
and can make what is a simple encoding problem into an unnecessarily complex 
and arcane set of rules, breaking the intuition. 

\paragraph{Dotted Vector Clocks}

When using causal histories, we have shown that knowing the last event 
could simplify comparison by simply checking if the last event is included in 
the causal history. 
This can still be done with vectors, if we keep track in which node the 
last event has been created. 
For example, when questioning if $x=[\mathbf{2},0,0] \caus y=[2,\mathbf{3},0]$, 
with boldface indicating the last event in each vector, we can simply test 
if $x[0] \leq y[0]$ ($\mathbf{2} \leq 2$) since we have marked that the last event in $x$ was created in node $a$, i.e., it corresponds to the first entry of the vector.
Since marking numbers in \textbf{bold} is not a very practical implementation, 
the last event is usually stored outside the vector (and sometimes called a \emph{dot}): 
e.g. $[2,\mathbf{2},0]$ can be represented as $[2,1,0]b_2$. 
Notice that now the vector represents the causal past of $b_2$, excluding the 
event itself. 

This strategy of decoupling the last event can also be captured graphically: 

\begin{center}
\begin{tabular}{lllll}
 $[2,\mathbf{2},0]$
 & $~\equiv$ & 
 $[2,1,0]b_2$
 & $~\equiv$ & 
\includegraphics{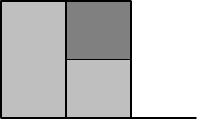}
\end{tabular}
\end{center}


\paragraph{Version Vectors}

In an important class of applications there is no need to register 
causality for all the events in a distributed computation. 
For instance, when modifying replicas of data, it often suffices 
to only register events that create new versions.
In this case, when thinking about causal histories, we only need to 
assign a new unique name to these relevant events.
Still, we need to propagate the causal histories when messages 
are propagated from one site to the other and the remaining rules
for comparing causal histories remain unchanged.

\begin{figure}[h!]
\begin{center}
\begin{xy}
\xymatrix{
  \mathit{Node \; A} & \ar@{.}[r] &
  {\bullet} \ar@{->}[r]^<{\{\mathbf{a_1}\}} & \circ \ar@{->}[rr]^<{\{\mathbf{a_1}\}} \ar@{->}[rd]
  & & {\bullet} \ar@{.}[rrr]^<{\{a_1,\mathbf{a_2}\}}  & & & \\
  \mathit{Node \; B} & \ar@{.}[r] & \bullet \ar@{->}[rr]^<{\{\mathbf{b_1}\}}
&  &
  {\bullet} \ar@{->}[rr]_<{\{a_1,b_1,\mathbf{b_2}\}}^<{\mathcal{M}} & & \circ \ar@{.}[rr]^<{\{a_1,b_1,\mathbf{b_2}\}} \ar@{->}[rd]
&  & \\
\mathit{Node \; C} & \ar@{.}[r] & \circ \ar@{->}[rrr]_<{\{\}}
& & &
{\circ} \ar@{->}[rr]_<{\{\}} &
  & {\circ} \ar@{.}[r]_<{\{a_1,b_1\mathbf{b_2}\}} 
& \\
  \mathit{Time} & \ar@{-->}[rrrrrrr] & & & & & & & \\
}
\end{xy} 
\end{center}
\caption{Run in a distributed system with three nodes, where only some events
are relevant: causal histories.}
\label{fig:vv:hb:example2}
\end{figure}
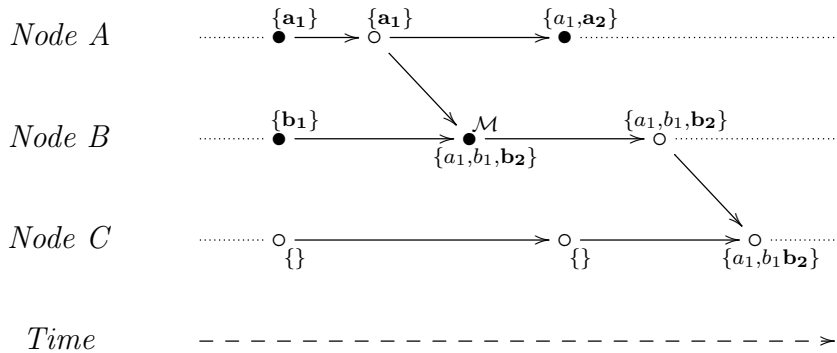

Figure \ref{fig:vv:hb:example2} presents the same example as before,
but with events not being registered for causality tracking 
denoted with $\circ$.
If the run represents the updates to replicas of a data object, we can
see that after node $a$ and $b$ are concurrently modified,
the state of replica $a$ is sent to replica $b$ (in a message).
When the message is received in node $b$, it is detected that two
concurrent updates have occurred, with histories $\{a_1\}$ and $\{b_1\}$,
as neither $a_1 \caus b_1$ nor $b_1 \caus a_1$.
In this case, a new version that merges the two updates is created (merge is denoted by the symbol $\mathcal{M}$), which 
requires creating a new unique name, leading to $\{a_1,b_1,b_2\}$.
When the state of replica $b$ is later propagated to replica $c$, 
as no concurrent update exist in replica $c$, no new version is created.

\begin{figure}
\begin{center}
\begin{xy}
\xymatrix{
  \mathit{Node \; A} & \ar@{.}[r] &
  {\bullet} \ar@{->}[r]^<{[\mathbf{1},0,0]} & \circ \ar@{->}[rr]^<{[\mathbf{1},0,0]} \ar@{->}[rd]
  & & {\bullet} \ar@{.}[rrr]^<{[\mathbf{2},0,0]}  & & & \\
  \mathit{Node \; B} & \ar@{.}[r] & \bullet \ar@{->}[rr]^<{[0,\mathbf{1},0]}
&  &
  {\bullet} \ar@{->}[rr]_<{[1,\mathbf{2},0]}^<{\mathcal{M}} & & \circ \ar@{.}[rr]^<{[1,\mathbf{2},0]} \ar@{->}[rd]
&  & \\
  \mathit{Node \; C} & \ar@{.}[r] & \circ \ar@{->}[rrr]_<{[0,0,0]}
& & &
  {\circ} \ar@{->}[rr]_<{[0,0,0]} &
  & \circ \ar@{.}[r]_<{[1,\mathbf{2},0]} 
& \\
  \mathit{Time} & \ar@{-->}[rrrrrrr] & & & & & & & \\
}
\end{xy} 
\end{center}
\caption{Run in a distributed system with three nodes, where only some events
are relevant: version vectors.}
\label{fig:vv:vv:example2}
\end{figure}
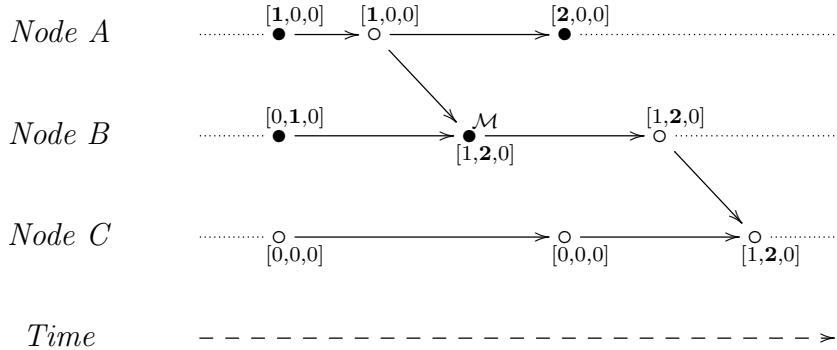

Again we can use vectors to compact the representation. The resulting 
representation is known as version vector and was created in 1983 \cite{ppr}, 
five years before vector clocks. Figure \ref{fig:vv:vv:example2} presents
the same example as before, represented with version vectors.

\begin{figure}[h!]
\begin{center}
\begin{xy}
\xymatrix{
  \mathit{Node \; A} & \ar@{.}[r] &
  {\bullet} \ar@{->}[r]^<{\{\mathbf{a_1}\}} & \circ \ar@{->}[rr]^<{\{\mathbf{a_1}\}} \ar@{->}[rd]
  & & {\bullet} \ar@{.}[rrr]^<{\{a_1,\mathbf{a_2}\}}  & & & \\
  \mathit{Node \; B} & \ar@{.}[r] & \bullet \ar@{->}[rr]^<{\{\mathbf{b_1}\}}
&  &
  {\circ} \ar@{->}[r]_<{\{\mathbf{a_1}\},\{\mathbf{b_1}\}} & {\bullet} \ar@{->}[r]^<{\{a_1,b_1,\mathbf{b_2}\}}_<{\mathcal{M}} & \circ \ar@{.}[rr]^(.25){\{a_1,b_1,\mathbf{b_2}\}} \ar@{->}[rd]
&  & \\
\mathit{Node \; C} & \ar@{.}[r] & \circ \ar@{->}[rrr]_<{\{\}}
& & &
{\circ} \ar@{->}[rr]_<{\{\}} &
  & {\circ} \ar@{.}[r]_<{\{a_1,b_1\mathbf{b_2}\}} 
& \\
  \mathit{Time} & \ar@{-->}[rrrrrrr] & & & & & & & \\
}
\end{xy} 
\end{center}
\caption{Run in a distributed system with three nodes, where only some events
are relevant and versions are not immediately merged: causal histories.}
\label{fig:vv:hb:example3}
\end{figure}
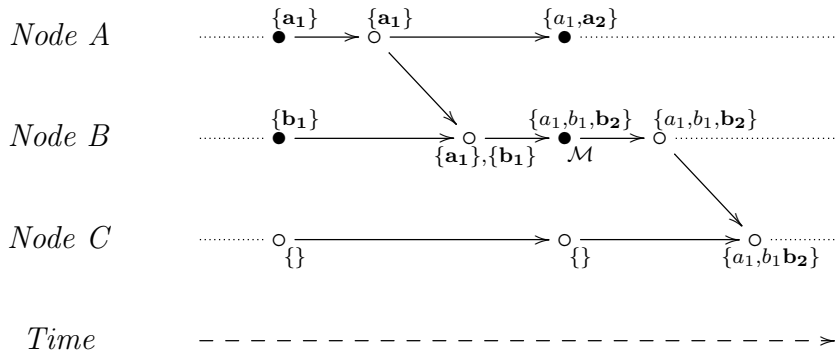

In some cases, when the state of one replica is propagated to the other 
replica, the two versions are kept by the system as conflicting versions.
For example, in Figure \ref{fig:vv:hb:example3}, when the message from
node $a$ is received in node $b$, the system keeps each causal history, 
$\{a_1\}$ and $\{b_1\}$, associated to the respective version.
The causal history associated with the node containing both version is
$\{a_1,b_1\}$, the union of the causal history of all versions. 
This approach allows to later check for causality 
relations between each version and other versions when 
merging the state of additional nodes.
The conflicting versions could also be merged, creating a new unique name,
as in the example.

\paragraph{Dotted Version Vectors}

One limitation of causality tracking by vectors is that one entry is needed for
each source of concurrency \cite{DBLP:journals/ipl/Charron-Bost91}. One can
expect a difference of several orders of magnitude from the number of nodes in
a data-center to the number of clients they handle. Vectors with one entry per
client, don't scale well when millions of clients are accessing the service
\cite{LOL}. Again, we can appeal to the foundation of causal histories to check
how to overcome this limitation. 

The basic requirement in causal histories is that each event is assigned a unique
identifier. There is no requirement that this unique identifier is created locally
nor that it is immediately created.
Thus, in systems where nodes can be divided in clients and servers and where 
clients communicate only with servers, it is possible to delay the creation of 
a new unique name until the client communicates with the server and
to use a unique name generated in the server.
The causal history associated with the new version is the union of the
causal history of the client and the newly assigned unique name.

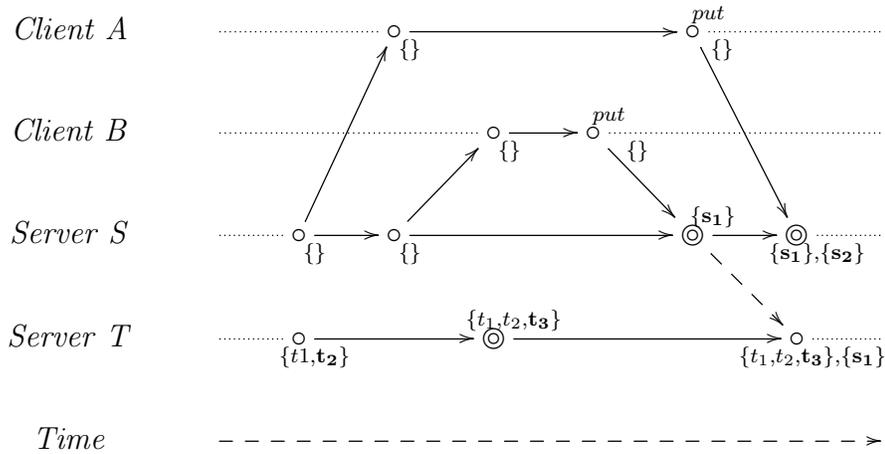
\begin{figure}[h!]
\begin{center}
\begin{xy}
\xymatrix{
  \mathit{Client \; A} & \ar@{.}[rr] & &
  \circ \ar@{->}[rrr]_<{\{\}} & & & \circ \ar@{.}[rr]^<{put}_(.15){\{\}} \ar@{->}[rdd] & &  \\
  \mathit{Client \; B} & \ar@{.}[rrr] & 
&  &
  \circ \ar@{->}[r]_<{\{\}} & \circ \ar@{.}[rrr]^<{put}_(.15){\{\}} \ar@{->}[rd] &
  &  &  \\
  \mathit{Server \; S} & \ar@{.}[r] & \circ \ar@{->}[r]_<{\{\}} \ar@{->}[uur]
  & \circ \ar@{->}[ur] \ar@{->}[rrr]_<{\{\}} & & 
  & {\varocircle} \ar@{->}[r]^<{\{\mathbf{s_1}\}} \ar@{-->}[rd]
  & \varocircle  \ar@{.}[r]_<{\{\mathbf{s_1}\},\{\mathbf{s_2}\}} &  \\
  \mathit{Server \; T} & \ar@{.}[r] & \circ \ar@{->}[rr]_<{\{t1,\mathbf{t_2}\}}
  & & \varocircle \ar@{->}[rrr]^<{\{t_1,t_2,\mathbf{t_3}\}}  &
  & & {\circ} \ar@{.}[r]_<{\{t_1,t_2,\mathbf{t_3}\},\{\mathbf{s_1}\}} 
  &  \\
  \mathit{Time} & \ar@{-->}[rrrrrrr] & & & & & & &  \\
}
\end{xy} 
\end{center}
\caption{Run in a distributed storage system: causal histories.}
\label{fig:dvv:hb:example}
\end{figure}

\begin{figure}[h!]
\begin{center}
\begin{xy}
\xymatrix{
  \mathit{Client \; A} & \ar@{.}[rr] & &
  \circ \ar@{->}[rrr]_<{[0,0]} & & & \circ \ar@{.}[rr]^<{put}_(.15){[0,0]} \ar@{->}[rdd] & &  \\
  \mathit{Client \; B} & \ar@{.}[rrr] & 
&  &
  \circ \ar@{->}[r]_<{[0,0]} & \circ \ar@{.}[rrr]^<{put}_(.15){[0,0]} \ar@{->}[rd] &
  &  &  \\
  \mathit{Server \; S} & \ar@{.}[r] & \circ \ar@{->}[r]_<{[0,0]} \ar@{->}[uur]
  & \circ \ar@{->}[ur] \ar@{->}[rrr]_<{[0,0]} & & 
  & {\varocircle} \ar@{->}[r]^<{[0,0]\mathbf{s_1}} \ar@{-->}[rd]
& \varocircle  \ar@{.}[r]_<{[0,0]\mathbf{s_1},[0,0]\mathbf{s_2}} &  \\
  \mathit{Server \; T} & \ar@{.}[r] & \circ \ar@{->}[rr]_<{[0,1]\mathbf{t_2}}
  & & \varocircle \ar@{->}[rrr]^<{[0,2]\mathbf{t_3}}  &
  & & {\circ} \ar@{.}[r]_<{[0,2]\mathbf{t_3},[0,0]\mathbf{s_1}} 
  &  \\
  \mathit{Time} & \ar@{-->}[rrrrrrr] & & & & & & &  \\
}
\end{xy} 
\end{center}
\caption{Run in a distributed storage system: dotted version vectors.}
\label{fig:dvv:vv:example}
\end{figure}
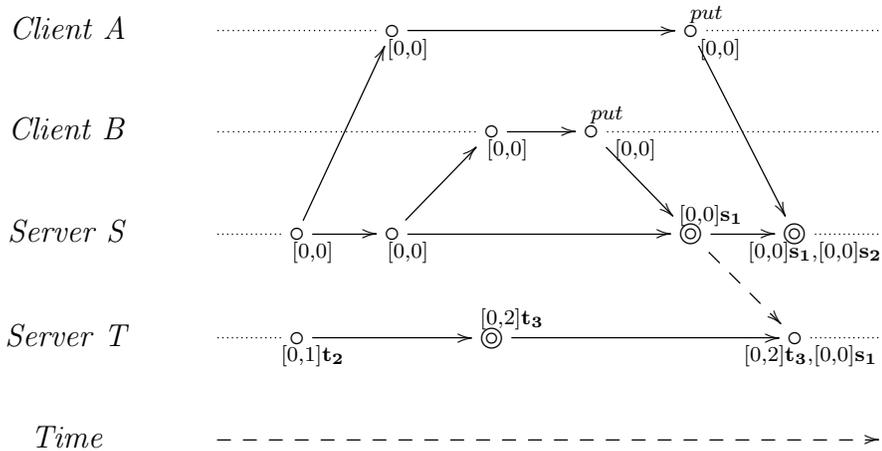

Figure \ref{fig:dvv:hb:example} shows an example where 
clients A and B concurrently update server S.
When client B first writes its version, a new unique name, $s_1$, is
created (in the figure we denote this action by the symbol $\varocircle$) which is merged with the causal history read by the client, $\{\}$,
leading to the causal history $\{\mathbf{s_1}\}$.
When client A later writes its version, the causal history assigned to this version
is the the causal history at the client, $\{\}$, merged with the new unique name, $s_2$,
leading to $\{\mathbf{s_2}\}$.
Using the normal rules for checking for concurrent updates, these
two versions are concurrent.
In the example, the system keeps both concurrent updates.
For simplicity we omitted interactions of server T with its own clients, but we can see that before receiving data from server S it had a single version that depicted three updates managed by server T, causal history $\{t_1,t_2,\mathbf{t_3}\}$, and after that it holds two concurrent versions. 

An important observation is that in each node, the union of the causal
histories of all versions includes all generated unique name until the last
known one: for example, in server S, after both clients send their
new versions, all unique names generated in S are known.
Thus, the causal past of any update can always be represented using
a compact vector representation, as it is the union of all versions
known at some server when the client read the object.
The combination of the causal past represented as a vector and 
the last event, kept outside the vector, is known as a 
\emph{dotted version vector} \cite{DBLP:conf/podc/PreguicaBAFG12,DBLP:conf/dais/AlmeidaBGPF14}. 
Figure \ref{fig:dvv:vv:example} shows the previous example using this
representation, that eventually becomes much more compact than causal histories as the system keeps running. 

The notion of compact vector representation is easier to understand under a graphical representation. An area with no \emph{hanging} columns or dots is compact, and can be efficiently encoded as a vector depicting column heights. 
Lets consider again the point in the execution after both clients updated S and it holds two concurrent versions. Graphically they can easily be checked to be concurrent since neither area includes the other: 

\begin{center}
\includegraphics{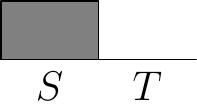}
$~~~\parallel$
\includegraphics{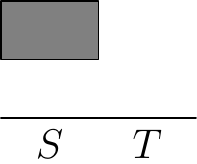}
\end{center}

A client that reads these versions with a \emph{get} operation, will get a causal context that reflects the join of these versions. This signifies that it was given access to the combination of the two concurrent causal histories and any update it does later should subsume these versions  

\begin{center}
\begin{tabular}{lll}
\includegraphics{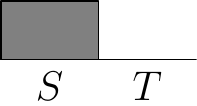}
$~~~\sqcup$
\includegraphics{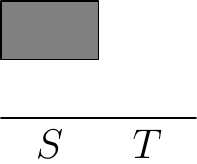}
 & $=$ & 
\includegraphics{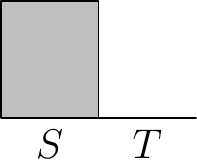}
\end{tabular}
\end{center}

if the client keeps server affinity and writes back to S with a \emph{put}, it gets assigned a new dot $\mathbf{s_3}$ there

\begin{center}
\includegraphics{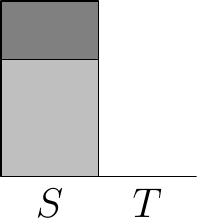}
\end{center}

and, due to area inclusion, this version will subsume the previous two in S. 

Alternatively, if the client could not contact S and wrote to T instead. It would get a different dot $\mathbf{t_4}$, but it would still lead to a representation that would subsume prior versions: 

\begin{center}
\includegraphics{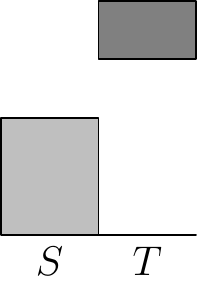}
\end{center}

In the condition expressed before (clients only communicate with
servers and a new update overwrites all versions previously read), 
which is common in key-value stores where multiple clients interact 
with storage nodes via a \emph{get/put} interface, the dotted version
vectors allow to track causality between the written versions 
with vectors of the size of the number of servers.


\paragraph{Identity Problems} 

We have seen that local creation of unique events requires each node to have exclusive access to a globally unique identity. We have glossed over this requirement by simply stating that node $i$ holds exclusive access to id $i$ and produces a sequence of unique events ids $i_1, i_2, \ldots$. Typically these identifiers are guaranteed to be unique by some external mechanism, like an administrator assigning different names to different nodes. 

In our graphic representation this can be captured by assigning control of different columns to different nodes. For instance, in a system with three replica nodes we can represent the initial state (still with no events registered) of each node's version vectors and identities:  

\begin{center}
\includegraphics{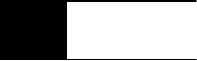}
\includegraphics{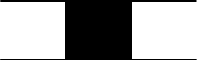}
\includegraphics{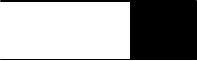}
\end{center}

Each node is only allowed to add events to the column it controls. For instance, the left node registered one event and the right node two events:

\begin{center}
\includegraphics{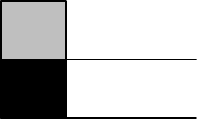}
\includegraphics{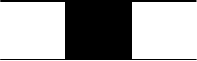}
\includegraphics{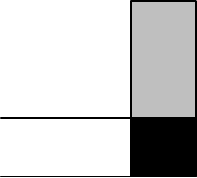}
\end{center}

If they all pair-wise join we converge to:

\begin{center}
\includegraphics{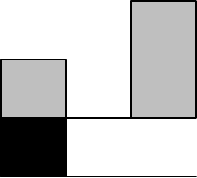}
\includegraphics{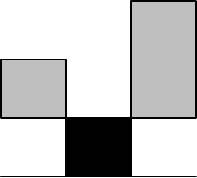}
\includegraphics{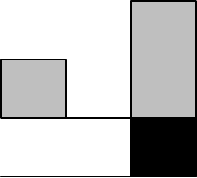}
\end{center}

\begin{figure}[h!]
\begin{center}
\begin{xy}
\xymatrix{
  \mathit{Node\; A} & \ar@{.}[r] &
  {\bullet} \ar@{->}[r]^<{1} & \bullet \ar@{->}[rr]^<{2} \ar@{->}[rd]
  & & {\bullet} \ar@{.}[rrr]^<{3}  & & & \\
  \mathit{Node \; B} & \ar@{.}[r] & \bullet \ar@{->}[rr]^<{1}
&  &
  {\bullet} \ar@{->}[rr]_<{3} & & \bullet \ar@{.}[rr]^<{4} \ar@{->}[rd]
&  & \\
  \mathit{Node\; C} & \ar@{.}[r] & \bullet \ar@{->}[rrr]_<{1}
& & &
  {\bullet} \ar@{->}[rr]_<{2} &
  & {\bullet} \ar@{.}[r]_<{5} 
& \\
  \mathit{Time} & \ar@{-->}[rrrrrrr] & & & & & & & \\
}
\end{xy} 
\end{center}
\caption{Run in a distributed system with three nodes: Lamport clocks.}
\label{fig:lc:example}
\end{figure}
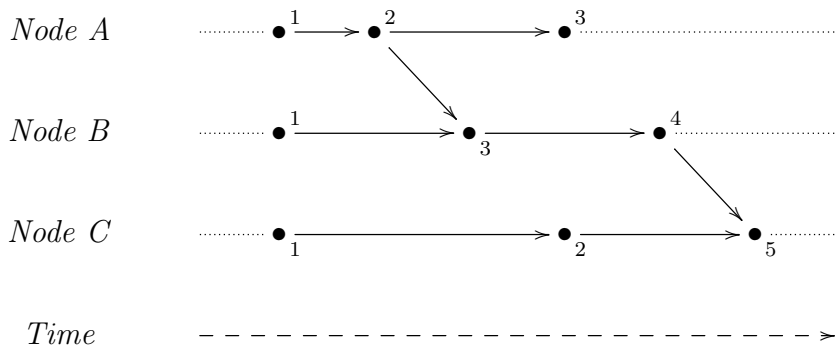

It is now appropriate to revisit the 1978 paper \cite{Lamport78} that defined causality in distributed systems. Although the paper does not introduce a mechanism that characterizes causality, it does introduce a total ordering that is consistent with causality, by way of scalar \emph{Lamport clocks}. Figure \ref{fig:lc:example} shows how our initial run is tagged with those clocks. Each node keeps a clock that starts at zero and is incremented on each event. Upon message reception the clocks are merged by maximum, and incremented to register the receive event.  

Using function $L$ to map a given event to its Lamport clock, we use the following expression to see how they are related to causality: $x \caus y$ implies that $L_x < L_y$. This total order, consistent with causality, might represent concurrent events as if they are related. 
However, those events that are causally ordered see that order preserved into the total order. 

Coming back to our graphical reasoning, we can observe that what happens is that all nodes are sharing the same identity. In the run on Figure \ref{fig:lc:example}, the system state after the execution is:

\begin{center}
\includegraphics{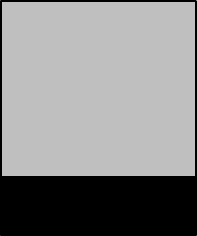}
\includegraphics{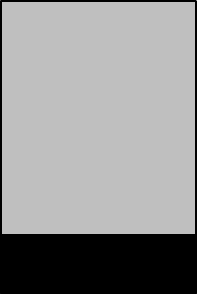}
\includegraphics{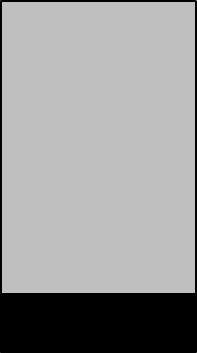}
\end{center}

Although we can see in the execution that the last event in A is concurrent to the last ones in B and C, the clock shows it to be ordered before both. 

Sharing identity has the obvious advantage of space economy since clocks no longer need to be linear on the number of active nodes \cite{DBLP:journals/ipl/Charron-Bost91}. This comes at the cost of no longer characterizing causality and missing detail on concurrent events. A middle ground is to partially share identities and provide a clock system that often detects concurrent events but sometimes might miss them. This is done with plausible clocks \cite{Torres-Rojas:1999:PCC:1035766.1035768}. The following configuration gives insight into a three node system with a modulo-2 plausible clock assignment, and thus with entry sharing among the left and right nodes:

\begin{center}
\includegraphics{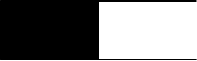}
\includegraphics{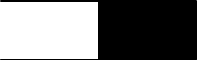}
\includegraphics{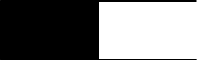}
\end{center}

Events that might occur concurrently in the left and right nodes will appear to be ordered. 

\paragraph{Interval Tree Clocks}

We have seen that sharing identities among active nodes can lead to non-characterization of causality and missing concurrency detection. The key to accurate causality tracking is to ensure that each node has access to a section of the identity line, that is exclusive to it, when it needs to register events. 
While some sharing is possible as long as exclusive sections are present \cite{Mostefaoui2017}, it is easier to keep the invariant that no sharing is allowed when addressing a dynamic solution to the management of identifiers. 

Interval tree clocks \cite{Almeida:2008:ITC:1496310.1496330} provide a
causality tracking approach that allows fine control on the set of active nodes
and their use of identifiers. The intuition, supported by the graphical
representation, is that a system comprising several  active nodes can be
derived by forking any active node, starting from a seed node that initially
holds the whole identifier space.  In the following example we derive a three
node replicated system by forking a seed replica and then forking again the
derived replica. Before forking, the seed replica has two events registered: 

\begin{center}
\includegraphics{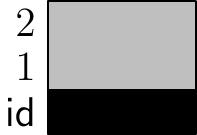}
\end{center}

After forking the seed, we have two active nodes, that contain the same causal past:

\begin{center}
\includegraphics{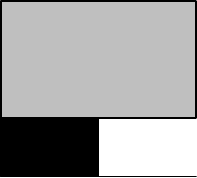}
\includegraphics{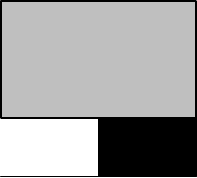}
\end{center}

Next we fork the rightmost one and locally register one new event in each, showing that nodes can only add areas inside the columns their current identity allows access to \--- i.e above the identity:

\begin{center}
\includegraphics{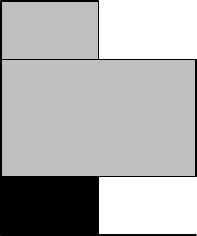}
\includegraphics{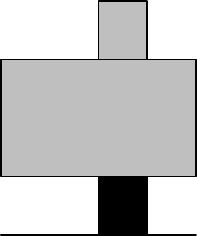}
\includegraphics{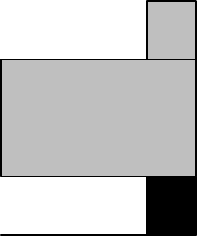}
\end{center}

Again, area inclusion allows a clear visual detection of concurrency. Interval tree clocks are implemented with a recursive encoding scheme that supports algorithmic comparison and correct event registration. We conclude by depicting a join of the rightmost into the leftmost node, showing that the event areas are merged and the identity components are added together: 

\begin{center}
\includegraphics{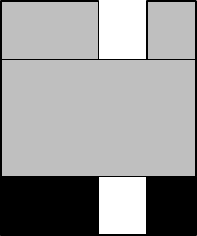}
\includegraphics{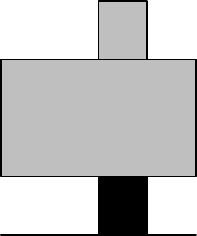}
\end{center}

\paragraph{Final remarks}

Tracking causality is important due to several reasons.
On one hand, not respecting causality can lead to strange behaviors for users
as reported by multiple authors \cite{cops,Ajoux15Challenges}.
On the other hand, tracking causality is important in the design of 
many distributed algorithms. 


The mechanisms for tracking causality and the rules used in these mechanisms 
are often seen as something complex \cite{VectorsEasy,VectorsHard} and their
presentation often lacks the necessary intuition of how they work. This perception of complexity has an impact on industry adoption. The initial Apache Cassandra design in 2009 \cite{CassandraITC} attempted to adopt causality tracking with vectors clocks and interval tree clocks, but complexity issues and the need for fast writes led the system to end up adopting wall clock timestamps, that can lead to lost updates. In another example \cite{RiakDVV}, the Riak key value store had state explosion problems, due to imperfect detection of causally subsumed replicas, that were only solved when the system was redesigned with dotted version vectors.

The most commonly used mechanisms for tracking causality, vector clocks and version vectors, are simply optimized representations of causal histories, which are easy to understand, and can be captured with a simple visual representation. 
By building on the notion of causal histories, we believe it is simple to
understand the intuition behind these mechanisms, to identify how they differ
and even possible optimizations. When confronted with an unfamiliar causality
tracking mechanism, or when trying to design a new system that requires it, we
urge the reader to fall back to two simple questions: (a) Which are the events
that need tracking? (b) How does the mechanism translate back to a simple
causal history? 

Without a simple mental image to guide us, errors and misconceptions become much more common. Sometimes, one only needs the right language. 

\paragraph{Acknowledgments} I would like to thank Nuno Pregui\c{c}a and Paulo S\'{e}rgio Almeida for their feedback on this work. 
This document was produced as part of the documentation for obtaining the \emph{agrega\c{c}\~{a}o} degree at Universidade do Minho. It is a modified and extended version of the dissemination article ``Why Logical Clocks are Easy'' of Baquero et. al. \cite{Baquero:2016:WLC:2907055.2890782}. 

\bibliographystyle{plain}
\bibliography{causality}

\end{document}